\documentclass[prd,nofootinbib,twocolumn,floatfix,preprintnumbers,showpacs]{revtex4}
\usepackage{graphicx}
\usepackage{dcolumn}
\usepackage{color}
\begin{document}
\newcommand{\be}{\begin{equation}}
\newcommand{\ee}{\end{equation}}
\newcommand{\bea}{\begin{eqnarray}}
\newcommand{\eea}{\end{eqnarray}}
\newcommand{\pp}{~~~.}
\newcommand{\vv}{~~~,}

\preprint{DFPD 05/A/11, DSF-02/2005, LAPTH-1088/05, IFIC/04-73}

\title{Do observations prove that cosmological neutrinos are
thermally distributed?}

\author{Alessandro Cuoco}
\affiliation{Dipartimento di Scienze Fisiche,
Universit\`{a} di Napoli {\it Federico II}, and INFN, Sezione di Napoli,
\\Complesso Universitario di Monte Sant'Angelo, Via Cintia, I-80126
Napoli, Italy}
\author{Julien Lesgourgues}
\affiliation{
LAPTH (CNRS-Universit\'{e} de Savoie), B.P.\ 110, F-74941
Annecy-le-Vieux Cedex, France,\\ and INFN, Sezione di Padova, Via
Marzolo 8, I-35131 Padova, Italy}
\author{Gianpiero Mangano}
\affiliation{Department of Physics, Syracuse University, Syracuse NY,
13244-1130, USA \\ and INFN, Sezione di Napoli and Dipartimento di
Scienze Fisiche, Universit\`{a} di Napoli {\it Federico II},
\\Complesso Universitario di Monte Sant'Angelo, Via Cintia, I-80126
Napoli, Italy}
\author{Sergio Pastor} \affiliation{Instituto de
F\'{\i}sica Corpuscular (CSIC-Universitat de Val\`{e}ncia),\\ Ed.\
Institutos de Investigaci\'{o}n, Apdo.\ 22085, E-46071 Valencia,
Spain}

\begin{abstract}
It is usually assumed that relic neutrinos possess a Fermi-Dirac
distribution, acquired during thermal equilibrium in the Early Universe.
However, various mechanisms could introduce strong distortions in this
distribution. We perform a Bayesian likelihood analysis including the first
moments of the three active neutrino distributions as free parameters, and
show that current cosmological observations of light element abundances,
Cosmic Microwave Background (CMB) anisotropies and Large Scale Structures
(LSS) are compatible with very large deviations from the standard picture.
We also calculate the bounds on non-thermal distortions which can be
expected from future observations, and stress that CMB and LSS data alone
will not be sensitive enough in order to distinguish between non-thermal
distortions in the neutrino sector and extra relativistic degrees of
freedom. This degeneracy could be removed by additional constraints from
primordial nucleosynthesis or independent neutrino mass scale measurements.
\end{abstract}

\maketitle

\section{Introduction}
\label{sec:intro}

The large amount of observations of the Cosmic Microwave Background (CMB)
which have been accumulated through forty years, since its discovery till
the most recent detailed CMB map provided by the Wilkinson Microwave
Anisotropy Probe (WMAP) Collaboration \cite{wmap}, shows that relic photons
are distributed according to a black body distribution to an incredibly
high accuracy. This fact is indeed one of the main cornerstones of the hot
Big Bang model, strongly supporting the idea that our Universe expanded
starting from high density and temperature conditions.

It is well known that in this framework we expect the Universe to be also
filled by a large amount of neutrinos, with present densities of the order
of 10$^2$ cm$^{-3}$ per flavor. These relic neutrinos decoupled from the
electromagnetic plasma quite early in time, when weak interaction rates
became slower than the Hubble rate for (photon) temperatures in the range
$2\div 4$ MeV, just before Big Bang Nucleosynthesis (BBN) took place.
Unfortunately the fact that this Cosmic Neutrino Background (C$\nu$B) has
today a very small kinetic energy, of the order of $10^{-4}$ eV, and that
neutrinos only interact via weak interactions, prevents us from any
possible direct detection of this background on the Earth (see {\it e.g.}
\cite{relicnu,relicnu2} for recent reviews).  Nevertheless, there are
several indirect ways to constrain the C$\nu$B by looking at cosmological
observables which are influenced either by the fact that neutrinos
contribute to the Universe expansion rate at all stages, or also via their
interactions with the electromagnetic plasma and baryons before their
decoupling. In this respect the most sensitive probe is represented by the
values of the light nuclide abundances produced during BBN. Actually, the
final yields of Deuterium, $^7$Li and in particular of $^4$He strongly
depend on the number of neutrino species as well as on their distribution
in phase space at about 1 MeV when the neutron to proton density ratio
freezes. In fact, since neutrinos were in chemical equilibrium with the
electromagnetic plasma till this epoch we know by equilibrium
thermodynamics that they were distributed according to a Fermi-Dirac
function, yet BBN can constrain exotic features like the value of their
chemical potential \cite{dolgovetal,abb,barger,cuoco}. Furthermore, a detailed
analysis carried out by several authors \cite{dolgov1,MMPP} shows that
neutrinos benefit to a small extent of the entropy released by
electron-positron pairs at their annihilation stage, and that this
phenomenon reheat neutrinos in a non-thermal way, introducing a small
distortion of the order of few percent which grows with the neutrino
comoving momentum.

Summarizing, neutrinos are likely to emerge from the BBN epoch as
decoupled species with an equilibrium distribution, a smaller
temperature with respect to photons by the approximate factor
$(4/11)^{1/3}$, possibly a small chemical potential to temperature
ratio $|\xi| \leq 0.1$\footnote{We point out that the bound on $\xi$
is shared by all neutrino species due to flavor oscillations
\cite{dolgovetal}, and that a much looser bound is obtained if we
allow for extra relativistic degrees of freedom contributing to the
Hubble rate during the BBN epoch (see {\it e.g.} \ \cite{cuoco}).} and
finally, tiny momentum dependent non-thermal features.

Once decoupled, neutrinos affect all key cosmological observables
which are governed by later stages of the evolution of the Universe
only via their coupling with gravity. A well known example is their
contribution to the total relativistic energy density, which affects
the value of the matter-radiation equality point, which in turn
influences the CMB anisotropy spectrum, in particular around the scale
of the first acoustic peak. Similarly, their number density and their
masses are key parameters in the small scale suppression of the power
spectrum of Large Scale Structures (LSS), smaller than $l_{nr}$, the
horizon when neutrinos become non-relativistic \cite{hu}
\be l_{nr}
\sim 38.5 \left(\frac{1~{\rm eV}}{m_\nu}\right)^{1/2} \omega_m^{-1/2}
~{\rm Mpc} \vv
\label{suppr}
\ee
where $\omega_m=\Omega_m h^2$, with $\Omega_m$ the fraction of the critical
density due to matter and $h$ the Hubble constant in units of $100$ km
s$^{-1}$ Mpc$^{-1}$. Several authors have investigated in details these
issues assuming a standard Fermi-Dirac distribution for neutrinos, see {\it
e.g.} \cite{barger,LP1,LP2,spergeletal,hannestad,hansenetal,melch}. Indications for primordial anisotropies in the neutrino distribution function have been pointed out in \cite{melch2}. In this
paper we consider the possibility that the relic neutrino distribution in
phase space may be sensibly different and address the following point: how
present (and future) cosmological observations can prove that indeed
neutrinos are thermally distributed?
The implications of a relic neutrino spectrum significantly different from a
Fermi-Dirac distribution have been also considered by several authors. For
instance, neutrinos could violate the Pauli exclusion principle and obey
Bose-Einstein statistics with important cosmological and astrophysical
implications \cite{BE1,BE2,Dolgov:2005mi}. Non-equilibrium neutrino spectra are also
produced in low-reheating scenarios
\cite{Giudice_reaheat,Adhya:2003tr,Hannestad_reheat}, from the decays of
massive neutrinos into relativistic products (see for instance
\cite{Hannestad_nudecay,Kaplinghat_decay}) or as the result of
active-sterile neutrino oscillations after decoupling (see
\cite{Abazajian:2004aj} and references therein). A full list of references
can be found in \cite{Dolgovrev}.

In order to answer the question raised above, we consider in detail
(Section \ref{sec:fnu}) a simple model where the out of equilibrium decay
of a light scalar produces non-thermal features in the neutrino spectra,
although our results are quite general and can be also applied to different
scenarios. We describe the current bounds on this model in Section
\ref{sec:current}, both from BBN and CMB+LSS. The expected sensitivity of
future cosmological data can be forecast if a fiducial cosmological model
is assumed, as described in Section  \ref{sec:future}. Finally we report
our concluding remarks in Section  \ref{sec:conclusions}.

\section{The phase-space distribution of relic neutrinos}
\label{sec:fnu}

As we already said, till temperatures of the order of MeV neutrinos are in
chemical equilibrium with the electromagnetic plasma, so that they keep an
equilibrium distribution. After weak interaction decoupling and in absence
of any other interaction process, the collisionless kinetic equations in a
Lemaitre-Friedman-Robertson-Walker universe state that, assuming for
simplicity instantaneous decoupling, the neutrino and antineutrino
distribution in phase-space is a Fermi-Dirac function. Therefore, in terms
of the comoving momentum $y=k a$, with $a$ the scale factor and neutrino
temperature $T_\nu$, one has
\be
df_{\alpha}(k,T_\nu) = \frac{1}{\pi^2} T_\nu^3 \, y^2 f_{\alpha}^{\rm th}(y)  \, dy
= \frac{1}{\pi^2} T_\nu^3 \, y^2 \frac{1}{e^y+1} \, dy
\ee
for each mass eigenstate. We assume zero chemical potential in the following.
Regardless of the specific case at hand, we can specify the 
neutrino distribution $df_\alpha(y)$ by the set of moments $Q_\alpha^{(n)}$
\be
Q_\alpha^{(n)} = \frac{1}{\pi^2}
\left(\frac{4}{11}\right)^{(3+n)/3}T^{3+n}
\int y^{2+n} f_\alpha(y) \, dy \vv
\label{momenta}
\ee
where $T$ is the photon temperature, and we have defined the
$Q_\alpha^{(n)}$ in terms of the standard value of the neutrino temperature
in the instantaneous decoupling limit $T_\nu= (4/11)^{1/3}T$. We assume in the following that neutrino distribution decays at large comoving momentum as $\exp(-y)$, so that the distribution admits moments of all orders. Actually, this is not a particularly severe constraint, since we expect that at very high $y$ the shape of the distribution is ruled by the behaviour imprinted by neutrino decoupling as hot relics at the MeV scale. 

If we denote by $P_m(y)$, $m$ being the degree of $P_m(y)$,
\be P_m(y) =  \sum_{k=0}^m c^{(m)}_k y^k \vv \ee 
the set of polynomials orthonormal with respect to the measure $y^2/(\exp(y)+1)$
\be \int_0^\infty dy \,\frac{y^2}{e^y + 1} P_n(y) P_m(y) = \delta_{nm} \vv \ee
it is easy to write the neutrino distribution in terms of its moments
\be df_\alpha(y) = \frac{y^2}{e^y+1} 
\sum_{m=0}^\infty F_{\alpha,m} P_m(y) \, dy \vv \label{series} \ee
where we have defined
\be F_{\alpha,m} = \sum_{k=0}^m c^{(m)}_k Q_\alpha^{(k)} T_\nu^{-k} \vv \ee
i.e. a linear combination of moments up to order $m$ with coefficients $c^{(m)}_k$.

For a Fermi-Dirac distribution all moments can be expressed in terms of the
number density $Q_\alpha^{(0)}$ or, equivalently, as functions of the only
independent parameter $T_\nu$. The first two moments, namely the neutrino
energy density defined during the relativistic regime and their number
density, enter all present analyses of cosmological observables. More
precisely, it is customary to define the effective number of neutrinos
\be
N_{\rm eff} = \frac{120}{7\pi^2} \left(\frac{11}{4}\right)^{4/3}
T^{-4}\sum_\alpha Q_\alpha^{(1)}
\vv
\label{neff}
\ee
where $N_{\rm eff}=3$ for a thermal distribution, while it gives $N_{\rm
eff}=3.04$ when including the small effect of neutrino heating during the
$e^+-e^-$ annihilation phase \cite{MMPP}.  Similarly, $Q_\alpha^{(0)}$ are
related to the present neutrino energy density normalized to the critical
density today
\be
\omega_{\nu} = \Omega_{\nu} h^2 = 0.058 \, \frac{m_0}{{\rm eV}}
\frac{11}{4}T^{-3} Q^{(0)}_\alpha \vv
\label{omnu}
\ee
where $m_0$ is the sum of the neutrino masses and we have assumed for
simplicity that the three neutrinos share the same distribution. 
Flavor neutrino oscillations are in fact very effective after BBN, since for
temperatures significantly lower than MeV the background effects of the medium are negligible, assuming the neutrino mixing parameters favoured by present experimental
results, see\cite{dolgovetal}. 
Therefore, the initial distortions on the flavor neutrino
spectra are fastly redistributed according to oscillations in vacuum.
Averaging over all oscillatory terms, it is easy to find the final
energy spectra of the three neutrino mass states. Here we take the
approximation of equal (non-thermal) distributions for all
neutrino mass states. In light of present sensitivity of data to neutrino
distribution, to be discussed in the following, this approximation is quite reasonable.

In the standard (thermal) case we recover the usual result
\be
\omega_{\nu} = \frac{m_0}{94.1 [93.2]\,{\rm eV}} \vv
\label{omnu:standard}
\ee
where the value between brackets takes into account the entropy release to
neutrinos from $e^+-e^-$ annihilations \cite{MMPP}. Actually, there are
more parameters which may affect the LSS power spectrum, such as the
individual neutrino masses, but their effect is negligible for a fixed
$m_0$.  

What about higher moments? In principle for an arbitrary distribution function, with the only requirement of an exponential decay for large comoving momentum, all moments may be relevant in the way this distribution is reconstructed by using the expansion (\ref{series}). In the following we will consider only the first two moments $\omega_{\nu}$ and $N_{\rm eff}$ as free and independent parameters, to be constrained using cosmological data. In fact, we start observing that in the limit of a thermal distribution, the value of $df_\alpha(y)$ is uniquely determined by $Q_\alpha^{(0)}$, see Eq. (\ref{series}). Adopting therefore a conservative approach one may think that if no extremely strong deviations from equilibrium are present, the expansion (\ref{series}) can be truncated for some $m$ without losing relevant pieces of information. The only way to decide how many parameters should be included in the analysis is by
studying the sensitivity of the available observational data to the distorsion 
of the neutrino distribution. As we will see, using all available data from CMB, LSS and possibly BBN, it is already very hard to get quite strong constraints on the first two moments, so that it appears that presently it is not worth comparing more complicated models with observations and introduce further parameters $Q_\alpha^{(n)}$. In fact more moments means more parameter degeneracies, so that the main conclusion of the analysis would remain unchanged. Of course, in case future observations would reach a higher sensitivity on neutrino distribution, it would be desirable to include higher order moments, such as the skewness or the kurtosis, related to $Q_\alpha^{(2)}$ and $Q_\alpha^{(3)}$, respectively, as free parameters in the analysis. 

We now turn to the main point: how the neutrino distributions may
acquire non-thermal shapes after the weak interaction freeze-out. A
possible source of such non standard contribution may arise from the decays
of unstable massive particles provided
\begin{itemize}
\item[i)] decays take place out of equilibrium, {\it i.e.} for
temperatures smaller than the decaying particle mass $M$. In fact, if
decays take place at equilibrium for $T \sim M$, the resulting neutrino
distribution would be still thermal, possibly with a different ratio of
neutrino-photon temperatures if the entropy is released in different
amounts to neutrinos and photons. Therefore, in this case the effect
consists at most in an overall constant factor in neutrino distribution,
changing the energy density in relativistic species.
\item[ii)] occur after weak interaction freeze-out.
\end{itemize}
The simplest scenario which we will be considering in details in the
following is the two body decay of a light ($M \leq 1$ MeV) neutral scalar
particle $\Phi$ whose dynamics is dictated by the following lagrangian
density
\be
{\cal L}= \frac{1}{2} \partial_\mu \Phi \partial^\mu \Phi -
\frac{1}{2} M^2 \Phi^2 - \frac{\lambda}{\sqrt{3}} \Phi \sum_i
\overline{\nu}_i\nu_i \label{lagr} + {\cal L}_{\rm int}\vv
\ee
with $i$ the flavor index and we have assumed a flavor independent
coupling $\lambda$ to neutrinos. With ${\cal L}_{\rm int}$ we denote the
interaction terms of $\Phi$ with all other fields except neutrinos and
responsible for the thermalization of the $\Phi$ quanta at high
temperatures. The $\Phi$ particles are assumed to decay in out of
equilibrium conditions before the photon last scattering epoch, so that the
produced neutrino burst directly influences the CMB anisotropy spectrum, as
well as the late LSS formation. Different scenarios can be envisaged, as
for example the case of unstable neutrinos $\nu_h$ decaying into a (pseudo)
scalar particle $\varphi$ and a lighter neutrino $\nu_l$,
\be \nu_h \rightarrow \varphi \, \nu_l \vv
\ee
or three-body decays. It is worth stressing that the aim of this paper is
to show how combining different cosmological observables we can constrain
the non-thermal contribution to the neutrino background. Despite of the
fact that we will be mainly dealing for definiteness with the out of
equilibrium $\Phi$ decay case, all our results are quite general and can be
applied to different scenarios as well.  Using Eq. (\ref{lagr}) we can
easily compute the decay rate
\be
\Gamma(\Phi \rightarrow \overline{\nu}\nu) =
\frac{\lambda^2 M}{8 \pi} \pp
\ee
If decays take place out of equilibrium for temperatures smaller than the
$\Phi$ mass it follows that the coupling $\lambda$ should be very small. In
the radiation dominated regime
\be
\lambda \sim 10^{-10} \frac{T_D/{\rm MeV}}{\sqrt{M/{\rm MeV}}} \vv
\label{lambda}
\ee
where $T_D$ is the temperature of the thermal part of the neutrino
background at decay. The coupling $\lambda$ also produces scattering
processes like $\Phi \nu \leftrightarrow \Phi \nu$ and crossed reactions,
which however receive contribution at order $\lambda^4$. It is easy to see
that for the small values implied by Eq. (\ref{lambda}) these processes
were never in equilibrium before $T_D$, so they cannot thermalize the
$\Phi$ particles with neutrinos after decoupling of weak interactions.
\begin{figure}
\includegraphics[width=.5\textwidth]{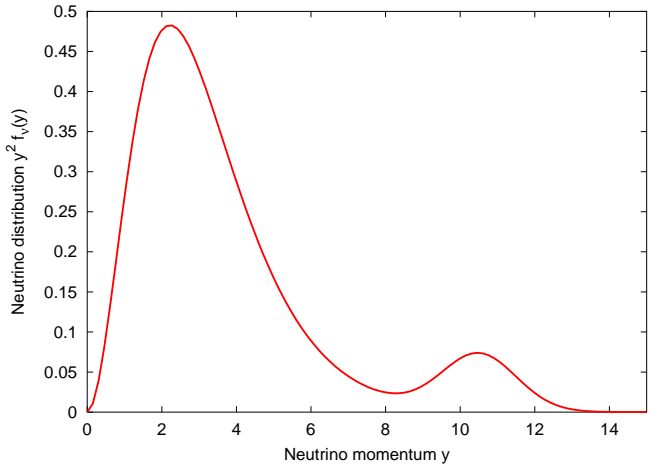}
\caption{\label{fig:NT} Differential number density of relic
neutrinos as a function of the comoving momentum for the non-thermal
spectrum in Equation (\ref{NTspec}). The parameters are $A=0.018$,
$y_*=10.5$ and $\sigma=1$, which corresponds to $N_{\rm eff} \simeq 4$.}
\end{figure}
When the $\Phi$ particles decay, the neutrino distribution gets an
additional contribution, which in the narrow width limit and in the
instantaneous decay approximation at $T_D$ corresponds to a peaked pulse at
$y_*=M/(2 T_D)$ so that
\bea
y^2 f(y) dy&=& y^2 \frac{1}{e^y+1} dy \nonumber \\ &+& \pi^2
\frac{A}{\sqrt{2 \pi \sigma^2}}
\exp \left[-\left(\frac{y-y_*}{2 \sigma^2}\right)^2\right] \vv
\label{NTspec}
\eea
where $\sigma \ll y_*$. Relaxing this assumption would be one of the many
ways to generalize our study. We show an example of this non-thermal
neutrino spectrum in Figure \ref{fig:NT}, where a second peak is clearly
seen around $y_*$. With this distribution, and accounting for tiny neutrino
heating effects from $e^+-e^-$ annihilations, the lower moments expressed
in terms of the parameters of Eqs. (\ref{neff}) and (\ref{omnu}) read
\begin{eqnarray}
\omega_{\nu} &=& \frac{m_0}{93.2 ~{\rm eV}} \left(1+ 0.99 \frac{2
\pi^2}{3
\zeta(3)}\,A \right) \vv \label{nonthomnu} \\
N_{\rm eff} &=& 3.04 \left(1+ 0.99 \frac{120}{7\pi^2}\,A y_* \right) \pp
\label{nonthneff}
\end{eqnarray}
The value of the constant $A$, which measures the number of neutrinos and
antineutrinos produced at decay can be expressed in terms of the total
number of $\Phi$ particles before decay in a comoving volume
\be
A = \frac{2}{3} \, \frac{n_\Phi}{T_\nu^3} \vv
\label{valuea}
\ee
with $n_\Phi$ the $\Phi$ number density. Since these particles are
decoupled from the thermal bath since at least the BBN epoch, we see that
the right hand side of Eq. (\ref{valuea}) can be evaluated at this early
epoch, so that the largest value of $A$ can be bound by BBN as a function
of the $\Phi$ mass and number density. We will discuss this issue in the
next Section along with the constraints on $A$ and $y_*$ arising from
present data on BBN, CMB and LSS.  Later we  will show in Section
\ref{sec:future} how future CMB and LSS data could improve these
constraints. Our results, which when expressed in terms of the parameters
$A$ and $y_*$ get a direct interpretation in the $\Phi$ particle decay
model described so far, nevertheless can be looked upon as quite general
and give the expected sensitivity that future data will likely have in
pointing out non-thermal features, if any. This sensitivity, as
unfortunately it is quite commonly the case, will be crucially depending on
the specific cosmological model which is assumed as a reference one, namely
on the number of parameters which enter the data fitting procedure. For
example, we will stress that if we enlarge this set of parameters allowing
for extra relativistic degrees of freedom in addition to ordinary photons
and neutrinos, a parameter degeneracy would make it very difficult to
distinguish between non-thermal corrections to the ordinary active neutrino
background and the presence of extra exotic light particles.

\section{Current bounds for a scenario with three non-thermal neutrinos}
\label{sec:current}

\subsection{Constraints from BBN}

The amount of light nuclei produced during BBN is rather sensitive to the
value of the Hubble parameter $H$ during that epoch, as well as to its time
dependence. In particular the $^4$He mass fraction $Y_p$ strongly depends
on the freeze-out temperature of weak processes which keep neutrons and
protons in chemical equilibrium. Changing the value of $H$ affects the
neutron to proton number density ratio at the onset of the BBN, which is
the key parameter entering the final value of $Y_p$ and more weakly the
Deuterium abundance. For a fixed baryon density parameter in the range
suggested by results of WMAP, $\omega_b=\Omega_b h^2= 0.023 \, {\pm}\, 0.001$
\cite{spergeletal} both nuclei yields are in fact monotonically increasing
functions of $H$. For recent reviews on BBN see
\cite{cuoco,serpico2004,cyburt2004}. This fact allows to severely constrain
any possible extra contribution to the total energy density due to other
species such as the decoupled massive scalar field $\Phi$ considered in the
present analysis.  Defining by $T_\Phi$ the corresponding temperature
parameter, the energy and number density of $\Phi$ particles can be written
as
\bea
\rho_\Phi &=& \frac{1}{2
\pi^2} T_\Phi^4 \int x^2 \left( x^2 + \frac{M^2}{T_\Phi^2}
\right)^{1/2} \frac{1}{e^x-1}\,dx \vv \label{edensity}\\
n_\Phi &=& \frac{1}{2 \pi^2} T_\Phi^3 \int x^2 \frac{1}{e^x-1}\, dx
= \frac{\zeta(3)}{\pi^2} T_\Phi^3 \vv
\label{ndensity}
\eea
where we have assumed, as from our discussion in the previous Section, that
the $\Phi$'s decoupled as hot relics well before the onset of BBN. We
mention that the effect of massive particles with mass of the order of $1
\div 10$ MeV which are instead coupled to neutrinos or photons during BBN
has been studied in details in \cite{SR}. The two Eqs. (\ref{edensity}) and
(\ref{ndensity}) can be combined getting
\bea
&& \rho_\Phi = T_\nu^4 \frac{1}{2 \zeta(3)}
\left(\frac{n_\Phi}{T_\nu^3} \right) {\times} \label{edensity2} \\
&&\int x^2 \left[ \left(\frac{\pi^2 n_\Phi}{\zeta(3) T_\nu^3}\right)^{2/3}
x^2 + \frac{M^2}{T_\nu^2} \right]^{1/2}
\frac{1}{e^x-1} \, dx \vv \nonumber
\eea
so that the upper bound on $\rho_\Phi$ from light nuclei abundances
translates into an upper limit to $n_\Phi /T_\nu^3$ and so to $A$, see Eq.
(\ref{valuea}), as a function of $M$. To get this bound we use a standard
likelihood procedure, as discussed {\it e.g.} in \cite{cuoco}, and the BBN
numerical code described in details in \cite{serpico2004}. We use the WMAP
prior on $\omega_b$ mentioned before and the experimental result of
\cite{kirkmanetal2003} for D/H
\begin{equation}
{\rm D/H}=(2.78^{+0.44}_{-0.38}) {\times} 10^{-5} \,\,\,.
\label{deut}
\end{equation}
We notice that, allowing for non-thermal features in neutrino distributions
may in principle change the preferred value of $\omega_b$ and the
corresponding uncertainty from a CMB data analysis. However, we will show
in the following that the effect on the baryon density parameter is
completely negligible and a likelihood analysis provides the same value of
a standard $\Lambda$CDM model, as considered in \cite{spergeletal}. It is
therefore consistent to adopt this result in our BBN study.

The $^4$He mass fraction $Y_p$ obtained by extrapolating to zero the
metallicity measurements performed in dwarf irregular and blue compact
galaxies is still controversial and possibly affected by systematics. There
are two different determinations \cite{fieldsolive1998,izotovthuan2004}
which are only compatible by invoking the large systematic uncertainty
quoted in \cite{fieldsolive1998}
\begin{eqnarray}
Y_p &=& 0.238 \, {\pm}\, (0.002)_{\rm stat}\, {\pm}\, (0.005)_{\rm sys} \,\,\,,\\
Y_p &=& 0.2421 \, {\pm}\, (0.0021)_{\rm stat}\,\,\,.
\end{eqnarray}
In the present analysis we consider a conservative estimate for the
experimental $^4$He abundance obtained by using the results of
\cite{oliveskillman2004}
\begin{equation}
Y_p^{\rm exp}=0.245 \, {\pm} \, 0.007 \,\,\,. \label{exphe}
\end{equation}
Finally, we do not use the $^7$Li abundance, since the experimental
estimates of this nuclide may be affected by a large depletion mechanism
with respect to the primordial value, see {\it e.g.}
\cite{cyburt2004,serpico2004}.
\begin{figure}
\includegraphics[width=.5\textwidth]{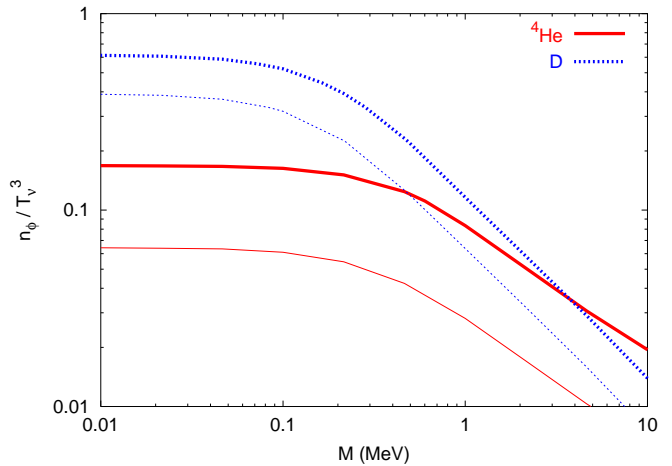}
\caption{\label{fig:BBNhe} The 1 $\sigma$ (thin lines) and 2 $\sigma$ (thick
lines) BBN bounds on the $\Phi$ number density (normalized to $T_\nu^3$)
versus mass $M$ in MeV. The regions above the contours would be in
disagreement with the observed primordial abundances of $^4$He or D.}
\end{figure}

Our results are summarized in Figure \ref{fig:BBNhe}, where we show the
1$\sigma$ and 2$\sigma$ bounds on $n_\phi/T_\nu^3$ obtained using the
$^4$He mass fraction and D/H, assuming that the $\Phi$'s decay after BBN.
As expected the most stringent constraint is due to $Y_p$, since this
parameter is very sensitive to any change of the value of the Hubble
parameter.  For $\Phi$ masses as large as 0.1 MeV, the scalars are
relativistic during the neutron to proton density freeze-out, so we get an
almost mass independent bound from $Y_p$ of the order of
$n_\phi/T_\nu^3\leq 0.17$ (or $A\leq 0.1$ at 2$\sigma$). This corresponds
to a limit on the non-thermal component of the neutrino number density of
the order $54\,\%$ of the thermal neutrino contribution. On the other hand,
for larger masses the upper bound on $n_\phi/T_\nu^3$ decreases
approximatively as $1/M$ for $M$ larger than few MeV, see Eq.
(\ref{edensity2}). For this mass range, the BBN bound from the $D$
abundance actually becomes slightly more stringent than that from $^4$He.
Note that the BBN bounds still allow a large contribution of the decay
products to the relativistic neutrino energy density $ N_{\rm eff}$. This
value depends quite crucially on the ratio $M/T_D$, which measures how far
from equilibrium conditions the $\Phi$ decays take place. For values of $M$
smaller than 0.1 MeV we get
\be N_{\rm eff} \leq 3.04 \left(1+ 0.1 \frac{M}{T_D} \right) \vv
\label{nonthneffbis}
\ee
where $T_D \leq 0.01$ MeV, since we assumed that the $\Phi$ decays take
place after BBN. If $T_D \sim 0.1 M$ or smaller the non-thermal component
can give a large or even dominant contribution to $ N_{\rm eff}$.  For
higher values $M\geq$ 1 MeV, our simple scenario becomes unlikely because
it requires the $\Phi$ decays to take place after BBN, and therefore one
needs a very large ratio $M/T_D \geq 10^2$.  However, in this case the
contribution to $N_{\rm eff}$ can still be very large, despite of the fact
that the $\Phi$ number density rapidly decreases with increasing values of
$M$. In fact, the upper bound on $N_{\rm eff}$ scales as $1/T_D$ in this
regime, see Eq.  (\ref{nonthneff}).

\subsection{Adding constraints from cosmological perturbations (CMB and LSS)}
\label{sec:current_bounds}

\begin{figure*}[t]
\includegraphics[width=.95\textwidth]{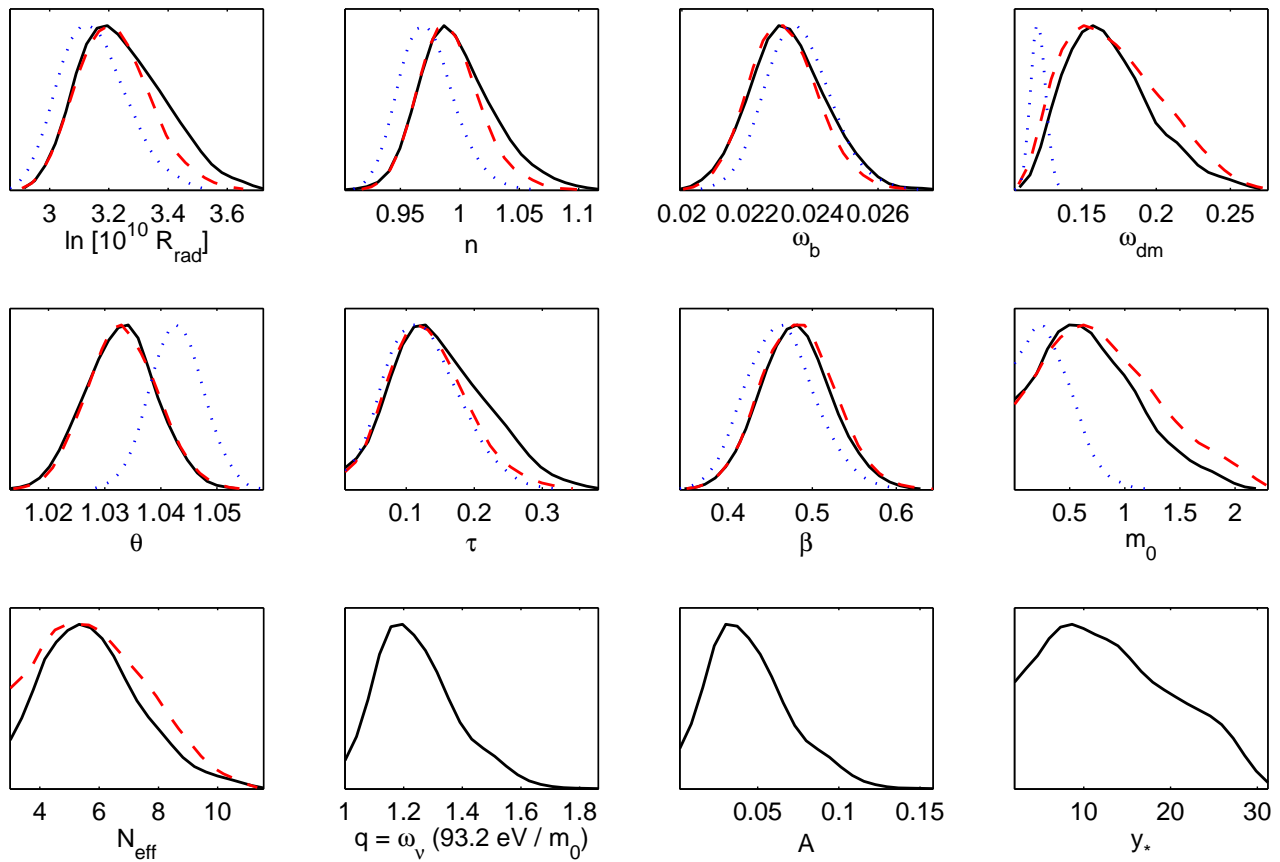}
\caption{Marginalized likelihood for the parameters of each model:
$\Lambda$CDM+NT (black, solid), $\Lambda$CDM+R (red, dashed), or
$\Lambda$CDM (blue, dotted). The independent basis parameters are included
in the first ten plots only. The last two plots show the related parameters
($A$, $y_*$) for the non-thermal model.
\label{1D_likelihoods}}
\end{figure*}
\begin{figure}
\includegraphics[width=.45\textwidth]{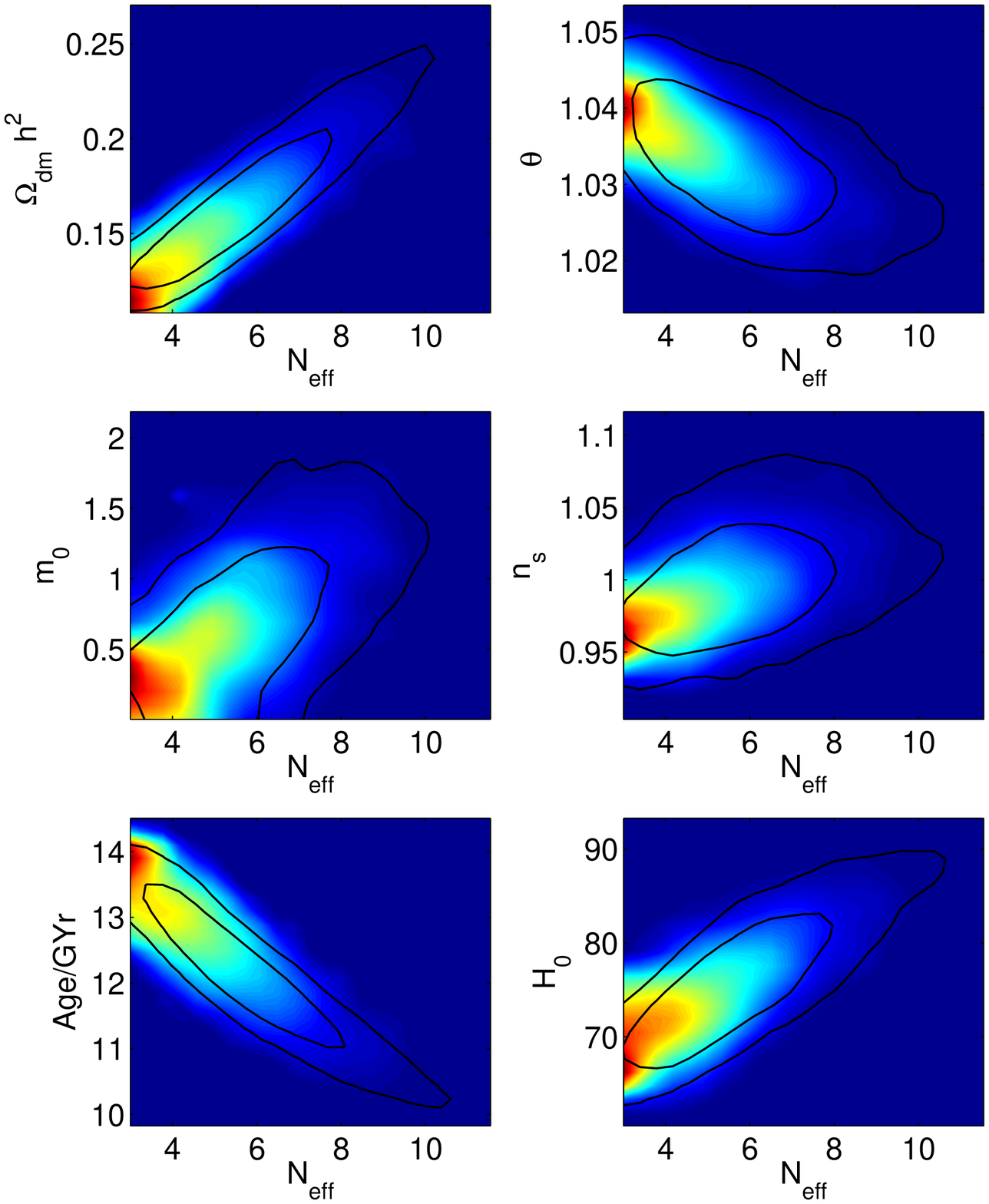}
\caption{Two-dimensional likelihoods illustrating the main degeneracy
affecting the $\Lambda$CDM+NT model (for the $\Lambda$CDM+R model the plots
are extremely similar). This degeneracy correlates $N_{\rm eff}$ mainly
with $\omega_{\rm dm}$ (top left) and $m_0$ (middle left), and to a smaller
extent with $\theta$ (top right) and $n_s$ (middle right). We also show how
$N_{\rm eff}$ is correlated with two related parameters: the age of the
Universe (bottom left) and the Hubble parameter (bottom right). The solid
curves refer to the 1-$\sigma$ and 2-$\sigma$ contours of the marginalized
likelihood. For comparison, the average sample likelihood is shown in color
levels. \label{2D_likelihoods}}
\end{figure}
We use the public code\footnote{CAMB Code homepage {\tt http://camb.info/}}
{\sc camb}~\cite{camb} in order to compute the theoretical prediction for
the $C_l$ coefficients of the temperature and polarization power spectra of
CMB, as well as the matter spectra $P(k)$. We explicitly implement into the
code the expression of the non-thermal phase-space distribution function,
in the same fashion as massive neutrino chemical potentials were added to
{\sc cmbfast}~\cite{cmbfast} in \cite{LP}. We focus on three models: the
minimal $\Lambda$CDM model including three massive neutrinos (with equal
mass); our extended model with three non-thermal massive neutrinos
($\Lambda$CDM+NT) and finally, for comparison a model with three ordinary
massive thermal neutrinos plus extra relativistic degrees of freedom
($\Lambda$CDM+R), already analyzed in \cite{HR,LP2} (using a grid-based
method and a smaller dataset). We compute the Bayesian likelihood of each
cosmological parameter with a Monte Carlo Markov Chain method, using the
public code {\sc cosmomc}~\cite{cosmomc} with option MPI on the {\sc
majorana} cluster at Naples University.
%
%
The likelihood of each model is found using three groups of data sets, as
implemented in the public version of {\sc cosmomc}: (i) CMB data:
WMAP~\cite{spergeletal}, VSA~\cite{VSA}, CBI~\cite{CBI},
ACBAR~\cite{ACBAR}; (ii) LSS data: 32 correlated points from 2dFGRS (up to
$k_{\rm max} = 0.1\, h$ Mpc$^{-1}$)~\cite{Peacock,2dFGRS}, 14 uncorrelated
points from SDSS~\cite{SDSS} (up to the same scale); and (iii) SNIa data
from Riess et al.~\cite{Riess2004}. For SDSS, we compare directly the data
with the real space linear matter power spectrum multiplied by a
scale-independent bias with flat prior, which we marginalize out. For 2dF,
we include a prior on the bias: the linear matter power spectrum (at
redshift $z=0$) is rescaled by a redshift-space correction at $z=0.17$ and
a bias factor~$b$
\begin{equation}
P(k)|^{\rm red.~sp.}_{z=0.17}=b^2 \left( 1+\frac{2}{3}
\beta_\mathrm{eff} + \frac{1}{5} \beta_\mathrm{eff}^2 \right)
P(k)|^{\rm real~sp.}_{z=0}
\end{equation}
with $\beta_\mathrm{eff}=0.85 \beta$ and $b=\Omega_m^{0.6}/\beta$ (see
\cite{WMAPverde} and references therein). For the redshift-space distortion
factor $\beta$ we adopt the prior $\beta=0.43{\pm}0.08$ \cite{Peacock}. We do
not include any data from Lyman-$\alpha$ forests\footnote{The non-thermal
model includes massive neutrinos which are individually more energetic and
with a higher average momentum than their thermal counterpart. In presence
of such a fluid, the non-linear small-scale evolution might differ
significantly from that of ordinary $\Lambda$CDM models. Therefore, we
believe that data from Lyman-$\alpha$ forests are inappropriate for a
conservative analysis. We are aware that the bias constraint does rely on
observations in the mildly non-linear range $0.2 < k < 0.4\, h$ Mpc$^{-1}$,
and we assume that on these scales the non-thermal neutrinos do not
generate unusually large non-linear corrections which would invalidate our
bias prior.}. Finally, we impose the gaussian prior $h=0.72 {\pm} 0.08$
(1$\sigma$) from the HST Key Project \cite{HST}, and for the non-thermal
model we add the BBN prior described in the previous Section: $A < 0.1$ at
the 2-$\sigma$ level (the prior is implemented as a half-gaussian centered
at $A=0$ and with a standard deviation of 0.05).  As we already said, a
remarkable feature of the $\Lambda$CDM+NT model is that BBN does {\it not}
impose an upper limit on $N_{\rm eff}$ at the CMB epoch, so the results
shown for comparison in the $\Lambda$CDM+R case will {\it not} include an
$N_{\rm eff}$ prior, and will not always be in agreement with predictions
from standard BBN. For the $\Lambda$CDM model, the basis of cosmological
parameters used by the Markov Chain algorithm is as follows: $^{1)}$ the
overall normalization parameter $\ln[10^{10} {\cal R}_{\rm rad}]$ where
${\cal R}_{\rm rad}$ is the curvature perturbation in the radiation era;
$^{2)}$ the primordial spectrum tilt $n_s$; $^{3)}$ the baryon density
$\omega_b=\Omega_b h^2$; $^{4)}$ the (cold+hot) dark matter density
$\omega_{\rm dm}=\Omega_{\rm dm} h^2$; $^{5)}$ $\theta$, the ratio of the
sound horizon to the angular diameter distance multiplied by 100; $^{6)}$
the optical depth to reionization $\tau$; $^{7)}$ the 2dF redshift-space
distortion factor $\beta$; $^{8)}$ the total mass $m_0$ of the three
neutrinos. In addition, the $\Lambda$CDM+R model includes: $^{9)}$ the
effective neutrino number $N_{\rm eff}=3.04+\Delta N_{\rm rel.}$. Finally,
the $\Lambda$CDM+NT model includes: $^{9)}$ the statistical moment
$Q_{\alpha}^{(1)}$ of the three non-thermal neutrinos, or equivalently,
their relativistic energy density in units of that of a standard massless
neutrino $N_{\rm eff}\geq3.04$, see Eq. (\ref{neff}); $^{10)}$ their
statistical moment $Q_{\alpha}^{(0)}$, or equivalently, the dimensionless
parameter $q \equiv \omega_{\nu} (93.2\,\mathrm{eV} / m_0) \geq 1$, see
Eqs. (\ref{omnu}) and (\ref{omnu:standard}).  The best value of the
effective $\chi^2$ for the three models is shown in the first line of Table
\ref{ranges}. The $\Lambda$CDM+R and $\Lambda$CDM+NT do not improve
significantly the $\chi^2$ despite of their larger parameter space (for a
precise statement on the goodness--of--fit of the various models, one
should carry a full ``evidence'' calculation as in~\cite{evidence}). On
Figure \ref{1D_likelihoods}, we plot the marginalized likelihood for each
parameter and model. The first ten plots correspond to basis (independent)
parameters, while the last two refer to the derived parameters ($A$, $y_*$)
for the non-thermal case, see Eqs. (\ref{nonthomnu}) and (\ref{nonthneff}).
The preferred values and 1-$\sigma$ errors for each parameter are shown in
Table \ref{ranges}.
\begin{table}[t]
\begin{tabular}{lccc}
\hline \hline
&$\Lambda$CDM&$\Lambda$CDM+R&$\Lambda$CDM+NT
\\
\hline
$\chi^2_{\rm min}$&1688.2&1688.0&1688.0
\\
\hline
$\ln[10^{10} {\cal R}_{\rm rad}]$&$3.2{\pm}0.1$&$3.2{\pm}0.1$&$3.2{\pm}0.1$\\
$n_s$&$0.97{\pm}0.02$&$0.99{\pm}0.03$&$1.00{\pm}0.03$\\
$\omega_b$&$0.0235{\pm}0.0010$&$0.0231{\pm}0.0010$&$0.0233{\pm}0.0011$\\ $\omega_{\rm
dm}$&$0.121{\pm}0.005$&$0.17{\pm}0.03$&$0.17{\pm}0.03$\\
$\theta$&$1.043{\pm}0.005$&$1.033{\pm}0.006$&$1.033{\pm}0.006$\\
$\tau$&$0.13{\pm}0.05$&$0.13{\pm}0.06$&$0.15{\pm}0.07$\\
$\beta$&$0.46{\pm}0.04$&$0.48{\pm}0.04$&$0.48{\pm}0.04$\\ $m_0$
(eV)&$0.3{\pm}0.2$&$0.8{\pm}0.5$&$0.7{\pm}0.4$\\ $N_{\rm eff}$&3.04&$6{\pm}2$&$6{\pm}2$\\
$q$&1&1&$1.25{\pm}0.13$\\
\hline
$h$&$0.67{\pm}0.02$&$0.76{\pm}0.06$&$0.76{\pm}0.05$\\
Age (Gyr)&$13.8{\pm}0.2$&$12.1{\pm}0.9$&$12.1{\pm}0.8$\\
$\Omega_{\Lambda}$&$0.68{\pm}0.03$&$0.67{\pm}0.03$&$0.67{\pm}0.03$\\
$z_{re}$&$14{\pm}4$&$16{\pm}5$&$18{\pm}6$\\
$\sigma_8$&$0.76{\pm}0.06$&$0.77{\pm}0.07$&$0.77{\pm}0.07$\\
\hline
\hline
\end{tabular}
\caption{
Minimum value of the effective $\chi^2$ (defined as $-2\ln{\cal L}$,
where ${\cal
L}$ is the likelihood function) and 1$\sigma$ confidence limits for the
parameters of the three models under consideration. The first ten lines
correspond to our basis of independent parameters, while
the last five refer to related parameters.
\label{ranges}}
\end{table}

For the parameters $\ln[10^{10} {\cal R}_{\rm rad}]$, $n_s$, $\omega_b$,
$\tau$, $\beta$, the likelihoods are almost equal in the three cases. In
par\-ti\-cular, the preferred range for the baryon density in the
$\Lambda$CDM+NT case is still the one of the standard $\Lambda$CDM model,
as we already anticipated in Section II.  On the other hand, in the
$\Lambda$CDM+R and $\Lambda$CDM+NT cases, the $\theta$ distribution is
significantly shifted, and $\omega_{\rm dm}$ and $m_0$ have much wider
error bars. The existence of a degeneracy direction involving mainly
$\omega_{\rm dm}$, $m_0$ and $N_{\rm eff}$ was already pointed out in
\cite{HR,LP2} for the $\Lambda$CDM+R model, and is confirmed here for the
two models in which the radiation density near the time of decoupling is a
free parameter\footnote{This degeneracy is due to the freedom to increase
the total radiation, total matter and critical densities, while keeping
fixed the redshift of equality and $\Omega_{\Lambda}$. More details can be
found in~\cite{LP2}.}. In particular, values as large as $\omega_{\rm
dm}=0.23$, $m_0 = 1.5$~eV or $N_{\rm eff}=9$ are still allowed at the
2-$\sigma$ level by our set of data and priors (while in the standard
$\Lambda$CDM case we find $\omega_{\rm dm}<0.130$ and $m_0 < 0.7$~eV at
2-$\sigma$, in agreement e.g. with \cite{spergeletal}). Note however that
in the $\Lambda$CDM+R case, models with $N_{\rm eff} > 4$ are in
disagreement with standard BBN, while in the $\Lambda$CDM+NT case they are
not. The degeneracy is illustrated by the two-dimensional likelihood plots
of Figure \ref{2D_likelihoods}. Finally, we can notice that in the
$\Lambda$CDM+NT model the bound on the parameter $q \equiv \omega_{\nu}
(93.2\,\mathrm{eV} / m_0)$ comes essentially from the BBN prior $A<0.1$,
since the CMB and LSS data alone would be compatible with much larger
deviations from a thermal phase-space distribution (up to $A=1$ at
2-$\sigma$). As a final remark we notice that $y_*$ is poorly constrained,
and large values for it are still allowed. In the $\Phi$ decay scenario
this implies that these decays can take place in highly out of equilibrium
conditions, $T_D \ll M$, the only bound being instead on the scalar particle
number density at the BBN epoch.

\section{Future prospects: can we remove the degeneracies?}
\label{sec:future}

\subsection{Error forecast assuming no extra relativistic degrees of freedom}

\begin{table*}
\begin{tabular}{lcccccccccccr}
\hline \hline
name &$\ln[10^{10} {\cal R}_{\rm rad}]$ &$n_s$  &$\omega_b$ &$\omega_m$
&$\Omega_{\Lambda}$ &$\tau$ &$Y_{\rm He}$ &$b^2$ &$m_0$ (eV)  &$N_{\rm
eff}$ &$q$ &
\\
\hline fiducial values
&3.1 &0.96 &0.023 &0.14 &0.70 &0.11 &0.24 &1.0 &0.5 &4.0 &1.1 &
\\
\hline 1$\sigma$ error
&0.01 &0.009 &0.0003 &0.004 &0.02 &0.005 &0.02 &0.08 &0.3 &0.1 &0.7 &
(conservative)
\\
&0.006 &~0.003~ &~0.0001~ &~0.0007~ &~0.001~ &~0.003~ &~0.004~ &0 &~0.03~
&~0.02~ &0.08 & (ambitious)
\\
\hline \hline
\end{tabular}
\caption{Expected errors on the parameters of the $\Lambda$CDM+NT
model. The first line shows the fiducial values, {\it i.e.} the parameter
values assumed to represent the best fit to the data. The next lines give
the associated 1$\sigma$ errors for the ``conservative'' and ``ambitious''
sets of future experiments as described in the text.
\label{table:secIIIa} }
\end{table*}

We will now compute the error expected from future experiments on the
parameters of the $\Lambda$CDM+NT model. Of course, the results that will
be derived here are model--dependent, since we assume that the cosmological
evolution can be described with a given set of cosmological parameters. The
issue of generalization to larger classes of models will be considered in
the next Section. As usually done in this kind of analysis, we first assume
that future data will favor a particular $\Lambda$CDM+NT model that we call
the ``fiducial model'', with cosmological parameter values $x_i=\bar{x}_i$.
Then, we compute the Fisher matrix
\begin{equation}
F_{ij}= - \frac{\partial^2 \ln {\cal L}}{\partial x_i \partial x_j}
(\bar{x}_i) \vv
\end{equation}
which represents the curvature of the likelihood distribution ${\cal L}$
near its maximum for a given  set of future experiments with known
sensitivity. Inverting the Fisher matrix, we obtain the 1-$\sigma$ error on
each parameter, assuming that all other parameters are unknown
\begin{equation}
\frac{\Delta x_i}{x_i} = (F^{-1})_{ii}^{1/2} \pp
\end{equation}
The present analysis follows the lines of that in
\cite{Lesgourgues:2004ps}, and we refer the reader to this work for details
on the expression of the likelihood distribution, for the description of
future experimental sensitivities and  for references of seminal papers on
the subject. An analysis for massless neutrinos and aimed to study the sensitivity on $N_{\rm eff}$ of WMAP and Planck experiments has been also discussed in \cite{bowen}.
When computing numerically  the derivative of the likelihood
with respect to each cosmological  parameter, we choose the smallest
possible step size above the threshold set by the precision of {\sc
cmbfast}, and we check that  the final result is almost independent of the
exact step value. Our first ``conservative'' set of experiments consists in
the Planck satellite~\footnote{\tt
http://www.rssd.esa.int/index.php?project=PLANCK} (without lensing
information \cite{QE}) combined with the completed SDSS redshift
survey~\footnote{\tt http://www.sdss.org} with effective volume $V_{\rm
eff}=1\,h^{-3}$ Gpc$^3$ and a free bias. We assume that the SDSS data can be
compared with the predictions of linear perturbation theory up to the scale
$k_{\rm max}=0.1\,h$ Mpc$^{-1}$. The second ``ambitious'' set consists in
CMBPOL~\footnote{\tt http://universe.nasa.gov/program/inflation.html}
combined with a large survey with effective volume $V_{\rm
eff}=40\,h^{-3}$ Gpc$^3$, no bias uncertainty, and $k_{\rm
max}=0.2\,h$ Mpc$^{-1}$, which should mimic the properties of a large
cosmic shear survey like the LSST project~\footnote{\tt
http://www.lsst.org} (this estimate should be regarded as indicative only,
since for simplicity we compute the Fisher matrix from the matter power
spectrum instead of the spectrum of the projected gravitational potential).

The results are shown in Table \ref{table:secIIIa} for a fiducial
model comfortably allowed by current bounds, with $N_{\rm eff}=4$,
$q=1.1$ and $m_0=0.5$~eV, corresponding to non-thermal correction
parameters $A=0.018$ and $y_*=10.5$.  For this model, the neutrinos
enter the non-relativistic regime well after decoupling, and only
contribute to 0.5\% of the critical density today.  As a consequence,
the two parameters $q$ and $m_0$ are measurable only from the
free-streaming effect in the matter power spectrum, while all other
parameters have a clear signature in the CMB anisotropies. More
precisely, a combination of these two parameters (close to $q {\times}
m_0 \propto \omega_{\nu}$ at first approximation) controls the
amplitude of the free-streaming suppression on small scales, while the
orthogonal combination controls the details of the transition on
intermediate scales. Since LSS experiments do not have enough
resolution for constraining this last combination, we expect a large
degeneracy between $q$ and $m_0$, which is confirmed by the
diagonalization of the Fisher matrix. The direction $\Delta q =
(\Delta m_0/0.5~ \mathrm{eV})$ appears to be poorly constrained. As a
consequence, the errors quoted in Table \ref{table:secIIIa} for $m_0$
are larger (typically by a factor two) than those expected for thermal
neutrinos, and $q$ has the biggest relative error. Notice that we are
not applying in this case any prior on $q$, as was instead the case in
the previous Section, where this parameter was related to the $\Phi$
number density at the BBN epoch in the $\Phi$ to neutrino decay
scenario. Therefore, results in Table \ref{table:secIIIa} are
independent of any particular model that produces a distortion on the
relic neutrino spectrum.
The remarkable result of this Section is that in spite of this degeneracy,
future experiments can provide some insight on non-thermal features and
still measure the neutrino mass with good precision even when the thermal
assumption is relaxed. The error for all parameters but ($q$, $m_0$) are
close to the usual expectations based on thermal neutrinos, see
\cite{Lesgourgues:2004ps} for comparison. 
We also find a sensitivity on $^4$He from CMB of the same order of magnitude found in \cite{trotta} for Planck experiment. A small distortion of the first
moment described by $q \sim 1.16$ would be sufficient for a 2-$\sigma$
detection with CMBPOL+LSST. Notice that large distortions (close to the
current upper bounds) would lead to smaller errors than those quoted in
Table \ref{table:secIIIa} due to their additional direct effect on the CMB.\\
As a final remark we note that in the CMBPOL+LSST scenario the uncertainty on the $^4$He mass fraction would be reduced by a factor two with respect to present sensitivity from direct measurements of this nuclide via spectroscopic methods. In the framework of the scalar decay scenario considered in this paper it is worth seeing whether this result from CMB may further constrain the value of $q$ and/or $N_{\rm eff}$. 
The result of course depends on the fiducial model which is considered, namely
what is the value of $^4$He which future measurements will suggest to be the 
preferred one. We consider as an example a fiducial model predicting $Y_p = 0.245$, in order to compare with constraints presently available, see Eq. (\ref{exphe}),
with an error of 0.004 (see Table II, "ambitious" set of data). Even in this
optimistic scenario, comparing this determination with the results of the nucleosynthesis theoretical code of \cite{serpico2004} we find the bound $q< 0.16$, a factor two larger than what is already obtained by CMB/LSS data (0.08, see again Table II). Finally, we recall that in the particle decay scenario considered in the paper we cannot bound the value of $N_{\rm eff}$ at CMB using BBN, since we are assuming that decays take place after BBN and the value of $N_{\rm eff}$ can be quite large depending on the $M/T_D$ ratio. Much stronger constraints come in fact from CMB/LSS data alone, order 0.1 and 0.02 in the conservative and ambitious scenarios respectively, see Table II.

\subsection{Degeneracy between non-thermal corrections and extra relativistic
degrees of freedom}

The bounds derived in the previous Sections depend on the assumption
that the cosmological observations can be described within the
framework of the $\Lambda$CDM+NT model. In the future, we would be in
a position to make such an analysis if we had some independent
motivations for the introduction of non-thermal distortions in the
neutrino background. Otherwise, even if the best-fit $\Lambda$CDM+NT
model turned out to be much more likely than the best-fit $\Lambda$CDM,
many alternative models would compete with each other.

The $\Lambda$CDM+NT model has a natural
``competitor'', which is the previously discussed $\Lambda$CDM+R model,
with three standard massive neutrinos and extra relativistic degrees of
freedom. Indeed, in the two models the amplitude and scale of the
free-streaming effect can be tuned independently, which is not usually the case
(and which cannot be mimicked by the introduction of tilt running,
spatial curvature, tensor modes, isocurvature modes, etc.)

The results of Section~\ref{sec:current_bounds} suggest that at
least within current experimental sensitivities, the two models are
roughly equivalent. However, many theoretical arguments suggest that the
intrinsic differences between the two cases can lead to distinct observable
signatures. For instance, the non-thermal corrections boost the average
neutrino momentum, so at the level of the equations the free-streaming
effect is never strictly identical in the two cases.

We have investigated in more details the issue of a possible (at least
approximate) one-to-one correspondence between $\Lambda$CDM+NT and
$\Lambda$CDM+R models. We studied many examples and found a systematic
way to find the ``twin $\Lambda$CDM+R model'' of a given
$\Lambda$CDM+NT model:
\begin{enumerate}
\item for any $\Lambda$CDM+NT model with a small total mass $m_0<0.9$
eV, the conclusion is straightforward: the $\Lambda$CDM+R model sharing the
same values of $N_{\rm eff}$ and $\omega_{\nu}$ has identical CMB
anisotropy spectra, and a matter power spectrum differing by less than one
percent on all scales. Note that the ``twin $\Lambda$CDM+R model'' has a
different total neutrino mass than the original $\Lambda$CDM+NT one, given
by the simple relation ${m}_0^{(R)}= q \, m_0^{(NT)}$.
\item for any $\Lambda$CDM+NT model with larger neutrino mass -- such
that the neutrinos are not ultra--relativistic at decoupling -- the
situation is a bit more complicated: the ``twin $\Lambda$CDM+R model'' does
not share that same values of $N_{\rm eff}$ and $\omega_{\nu}$, because the
time of equality is not strictly the same for the two models. Therefore, it
is necessary to increase slightly $N_{\rm eff}$ and $\Omega_{\Lambda}$ in
the $\Lambda$CDM+R model in order to obtain the same position and amplitude
of the first peak in the CMB temperature spectrum. Finally, a simultaneous
increase of $N_{\rm eff}$, $\omega_{m}$ and $n_s$ makes it possible to
match the matter power spectra without changing the time of equality and
the shape of the CMB temperature spectrum.
\end{enumerate}
In both cases the differences are below the percent level for the matter
power spectrum, even less for the CMB anisotropy spectra. This is too small
for allowing any detection with the next generation of
experiments\footnote{This statement is easy to prove quantitatively by
repeating the Fisher matrix analysis of the previous section, including an
extra parameter describing the ``linear deformation'' of one model into
another. This parameter results to be far from detectable.}. Of course,
this conclusion applies to a pure CMB and LSS analysis where all parameters
are assumed to be unknown. Strong evidence for the $\Lambda$CDM+NT could
still arise from external data sets: for instance, in case of contradiction
(assuming standard neutrinos) between future CMB and BBN predictions for
the value of $N_{\rm eff}$; or between the neutrino mass scale measured by
cosmology and by beta decay experiments.

\section{Conclusions}
\label{sec:conclusions}

In this paper we have studied the possibility that the phase space
distribution of cosmological relic neutrinos may be non-thermal, and how
present observations can constrain any departure from a standard
Fermi-Dirac function. The latter is expected to hold to a very good
accuracy (percent level) in the standard scenario, where neutrinos last
interact at the freeze out of weak interactions for temperatures of the
order of MeV, and then simply free stream keeping an equilibrium
distribution. However, if neutrino evolution is influenced at some later
stage by some unknown and exotic phenomena, we expect that their
distribution could be modified. For example, if some decoupled particle
eventually decay into neutrinos in out of equilibrium conditions and after
weak interaction freezing, the neutrino distribution can have large
departures from a simple Fermi-Dirac function. We have investigated this
model in details, for the case of light ($M\leq 1$ MeV) decoupled scalar
particles decaying in neutrino/antineutrino pairs.

Neutrinos both influence the CMB spectrum and the clustering of structures
on large scales. Their distribution, in fact, enters the value of the
matter-radiation equality point, as well as for massive neutrinos the
matter power spectrum suppression at scales smaller than the free streaming
scale. The neutrino distribution can be always fully characterized by the
set of its moments $Q_\alpha^{(n)}$ which in principle may all affect the
cosmological evolution of perturbations. However, present data are only
sensitive to the first two distribution moments. Actually, the total
neutrino mass $m_0$ and the moment $Q_\alpha^{(1)}$ (corresponding to
$N_{\rm eff}$) are well constrained by the CMB and LSS data, while
$Q_\alpha^{(0)}$ (corresponding to the ratio $\omega_{\nu}/m_0$) is not.
This fact, along with the existence of a degeneracy among $\omega_{\rm
dm}$, $N_{\rm eff}$ and $m_0$, implies that it is presently impossible to
establish whether neutrinos are thermally distributed or not, even in the
most conservative scenario of no extra relativistic particles at CMB epoch
apart from photons and neutrinos.  In particular, in the framework of the
$\Phi$ decay scenario described in Section \ref{sec:fnu}, the most
stringent constraint on $Q_\alpha^{(0)}$ comes from BBN only, $A \leq 0.1$,
rather than from CMB and LSS.  The non-thermal contribution to the total
neutrino energy density can then be very large or even dominant, since it
is ruled by the parameter $y_*=M/(2 T_D)$ which is very poorly constrained
and can take very large values, $y_* \sim 10$, even for $A$ close to the
BBN upper bound.

We have also studied how these results may likely change when new data on
CMB and LSS will be available. Adopting a Fisher matrix analysis we find
that combining the sensitivity of the Planck experiment with the completed
SDSS redshift survey, the uncertainty on the neutrino first moment $N_{\rm
eff}$ would be reduced by almost an order of magnitude, but still the
neutrino number density will be affected by a large error, see for example
the results of Table \ref{table:secIIIa}. The sensitivity to non-thermal
features in the neutrino distribution highly improves only by considering a
survey with the properties of a shear survey like the LSST project, and
also the sensitivity of a CMB experiment like CMBPOL. Even in this case,
however, it may be quite involved to test the hypothesis of a non thermal
neutrino background since, as we stressed in Section \ref{sec:future}, a
purely thermal neutrino background with extra relativistic degrees of
freedom would provide an almost degenerate scenario, with the exception of
predicting a different energy density at the time of primordial
nucleosynthesis, and a different value for the neutrino mass scale. In the
future, evidence for non-thermal distorsions could possibly result from an
{\it apparent} contradiction between the values of $N_{\rm eff}$ and/or
$m_0$ probed by cosmological perturbations (assuming standard neutrinos)
and those indicated respectively by primordial nucleosynthesis and particle
physics experiments. In particular, an independent source of information on
$m_0$ would greatly help. Direct neutrino mass measurements like the
Tritium beta decay experiment KATRIN \cite{katrin} or experiments on
neutrinoless double beta decay \cite{doublebeta}, which are expected to
reach a sensitivity on the neutrino mass scale of the order of 0.1$\div$0.3
eV, combined with cosmological observations could reduce this degeneracy.

Despite of the fact that by Ockham's razor principle it is presently worth
assuming a purely thermal distribution for relic neutrinos, nevertheless we
think that it is remarkable that present knowledge of the C$\nu$B features
still allows for quite non standard and exotic physics which affected the
evolution of the Universe at late stages. As frequently happened in the
past, neutrinos can play a role in constraining such possibilities.

\section*{Acknowledgments}

This research was supported by a Spanish-Italian AI, the Spanish grant
BFM2002-00345, as well as CICYT-IN2P3 and CICYT-INFN agreements. SP was
supported by a Ram\'{o}n y Cajal contract of MEC.


\begin{thebibliography}{100}

\bibitem{wmap}
C.L.~Bennett et al.,
Astrophys.\ J.\ Suppl.\  {\bf 148}, 1 (2003) [astro-ph/0302207].

\bibitem{relicnu}
G.B.~Gelmini,
hep-ph/0412305.

\bibitem{relicnu2}
C.~Hagmann,
astro-ph/9905258.

\bibitem{dolgovetal}
A.D.~Dolgov et al.,
Nucl.\ Phys.\ B {\bf 632}, 363 (2002) [hep-ph/0201287].

\bibitem{abb}
K.~N.~Abazajian, J.~F.~Beacom and N.~F.~Bell,
Phys.\ Rev.\ D {\bf 66}, 013008 (2002) [astro-ph/0203442].

\bibitem{barger}
V.~Barger et al.,
Phys.\ Lett.\ B {\bf 566}, 8 (2003) [hep-ph/0305075].

\bibitem{cuoco}
A.~Cuoco et al.,
Int.\ J.\ Mod.\ Phys.\ A {\bf 19}, 4431 (2004) [astro-ph/0307213].

\bibitem{dolgov1}
A.D.~Dolgov, S.H.~Hansen and D.V.~Semikoz,
Nucl.\ Phys.\ B {\bf 503}, 426 (1997) [hep-ph/9703315].

\bibitem{MMPP}
G.~Mangano, G.~Miele, S.~Pastor and M.~Peloso,
Phys.\ Lett.\ B {\bf 534}, 8 (2002) [astro-ph/0111408].

\bibitem{hu}
W.~Hu, D.J.~Eisenstein and M.~Tegmark,
Phys.\ Rev.\ Lett.\  {\bf 80}, 5255 (1998) [astro-ph/9712057].

\bibitem{LP1}
P.~Crotty, J.~Lesgourgues and S.~Pastor,
Phys.\ Rev.\ D {\bf 67}, 123005 (2003) [astro-ph/0302337].

\bibitem{LP2}
P.~Crotty, J.~Lesgourgues and S.~Pastor,
Phys.\ Rev.\ D {\bf 69}, 123007 (2004) [hep-ph/0402049].

\bibitem{spergeletal}
D.N.~Spergel et al.,
Astrophys.\ J.\ Suppl.\  {\bf 148}, 175 (2003) [astro-ph/0302209].

\bibitem{hannestad}
S.~Hannestad,
hep-ph/0412181.

\bibitem{hansenetal}
S.H.~Hansen et al.,
Phys.\ Rev.\ D {\bf 65}, 023511 (2002) [astro-ph/0105385].

\bibitem{melch}
G.L.~Fogli et al.,
Phys.\ Rev.\ D {\bf 70}, 113003 (2004) [hep-ph/0408045].

\bibitem{melch2} R. Trotta and A. Melchiorri, astro-ph/0412066.

\bibitem{BE1}
L.~Cucurull, J.A.~Grifols, and R.~Toldr\`{a},
Astropart.\ Phys.\  {\bf 4}, 391 (1996) [astro-ph/9506040].

\bibitem{BE2}
A.D.~Dolgov and A.Yu.~Smirnov,
hep-ph/0501066.

\bibitem{Dolgov:2005mi}
A.D.~Dolgov, S.H.~Hansen and A.Yu.~Smirnov,
astro-ph/0503612.

\bibitem{Giudice_reaheat}
G.F.~Giudice et al.,
Phys.\ Rev.\ D {\bf 64}, 043512 (2001) [hep-ph/0012317].

\bibitem{Adhya:2003tr}
P.~Adhya, D.~R.~Chaudhuri and S.~Hannestad,
Phys.\ Rev.\ D {\bf 68}, 083519 (2003) [astro-ph/0309135].

\bibitem{Hannestad_reheat}
S.~Hannestad,
Phys.\ Rev.\ D {\bf 70}, 043506 (2004) [astro-ph/0403291].

\bibitem{Hannestad_nudecay}
S.~Hannestad,
Phys.\ Rev.\ D {\bf 59}, 125020 (1999) [astro-ph/9903475].

\bibitem{Kaplinghat_decay}
M.~Kaplinghat, R.E.~Lopez, S.~Dodelson and R.J.~Scherrer,
Phys.\ Rev.\ D {\bf 60}, 123508 (1999) [astro-ph/9907388].

\bibitem{Abazajian:2004aj}
K.~Abazajian, N.F.~Bell, G.M.~Fuller and Y.Y.~Wong,
astro-ph/0410175.

\bibitem{Dolgovrev}
A.D.~Dolgov,
Phys.\ Rept.\  {\bf 370}, 333 (2002) [hep-ph/0202122].

\bibitem{serpico2004}
P.D.~Serpico et al.,
JCAP {\bf 0412}, 010 (2004) [astro-ph/0408076].

\bibitem{cyburt2004}
R.H.~Cyburt,
Phys.\ Rev.\ D {\bf 70}, 023505 (2004) [astro-ph/0401091].

\bibitem{SR}
P.D.~Serpico and G.G.~Raffelt,
Phys.\ Rev.\ D {\bf 70}, 043526 (2004) [astro-ph/0403417].

\bibitem{kirkmanetal2003}
D.~Kirkman et al.,
Astrophys.\ J.\ Suppl.\  {\bf 149}, 1 (2003) [astro-ph/0302006].

\bibitem{fieldsolive1998}
B.D.~Fields and K.A.~Olive,
Astrophys.\ J.\ {\bf 506}, 177 (1998) [astro-ph/9803297].

\bibitem{izotovthuan2004}
Y.I.~Izotov and T.X.~Thuan,
Astrophys.\ J.\  {\bf 602}, 200 (2004) [astro-ph/0310421].

\bibitem{oliveskillman2004}
K.A.~Olive and E.D.~Skillman,
astro-ph/0405588.

\bibitem{camb}
A.~Lewis and A.~Challinor,
Phys.\ Rev.\ D {\bf 66}, 023531 (2002) [astro-ph/0203507].

\bibitem{cmbfast}
U.~Seljak and M.~Zaldarriaga,
Astrophys.\ J.\  {\bf 469}, 437 (1996) [astro-ph/9603033].

\bibitem{LP}
J.~Lesgourgues and S.~Pastor,
Phys.\ Rev.\ D {\bf 60}, 103521 (1999) [hep-ph/9904411].

\bibitem{HR}
S.~Hannestad and G.~Raffelt,
JCAP {\bf 0404}, 008 (2004) [hep-ph/0312154].

\bibitem{cosmomc}
A.~Lewis and S.~Bridle,
Phys.\ Rev.\ D {\bf 66}, 103511 (2002) [astro-ph/0205436].

\bibitem{VSA}
R.~Rebolo et al., Mon.\ Not.\ Roy.\ Astron.\ Soc.\ {\bf 353}, 747 (2004)
[astro-ph/0402466];
C.~Dickinson et al., Mon.\ Not.\ Roy.\ Astron.\ Soc.\ {\bf 353}, 747 (2004)
 [astro-ph/0402498].

\bibitem{CBI}
T.J.~Pearson et al.,
Astrophys.\ J.\  {\bf 591}, 556 (2003) [astro-ph/0205388];
J.L.~Sievers et al.,
Astrophys.\ J.\  {\bf 591}, 599 (2003) [astro-ph/0205387];
A.C.S.~Readhead et al.,
Astrophys.\ J.\  {\bf 609}, 498 (2004) [astro-ph/0402359].

\bibitem{ACBAR}
C.~Kuo et al.,
Astrophys.\ J.\  {\bf 600}, 32 (2004) [astro-ph/0212289];
J.H.~Goldstein et al.,
Astrophys.\ J.\  {\bf 599}, 773 (2003) [astro-ph/0212517].

\bibitem{Peacock}
J.A.~Peacock et al.,
Nature {\bf 410}, 169 (2001) [astro-ph/0103143].

\bibitem{2dFGRS}
W.J.~Percival et al.,
Mon.\ Not.\ Roy.\ Astron.\ Soc.\  {\bf 327}, 1297 (2001)
[astro-ph/0105252];
Mon.\ Not.\ Roy.\ Astron.\ Soc.\ {\bf 337}, 1068 (2002) [astro-ph/0206256].

\bibitem{SDSS}
M.~Tegmark et al.,
Astrophys.\ J.\  {\bf 606}, 702 (2004) [astro-ph/0310725].

\bibitem{Riess2004}
A.G.~Riess et al.,
Astrophys.\ J.\  {\bf 607}, 665 (2004) [astro-ph/0402512].

\bibitem{WMAPverde}
L.~Verde et al.,
Astrophys.\ J.\ Suppl.\  {\bf 148}, 195 (2003) [astro-ph/0302218].

\bibitem{HST}
W.L.~Freedman et al.,
Astrophys.\ J.\  {\bf 553}, 47 (2001) [astro-ph/0012376].

\bibitem{evidence}
M.~Beltr\'{a}n, J.~Garc\'{\i}a-Bellido, J.~Lesgourgues,
A.R.~Liddle and A.~Slosar,
astro-ph/0501477.

\bibitem{Lesgourgues:2004ps}
J.~Lesgourgues, S.~Pastor and L.~Perotto,
Phys.\ Rev.\ D {\bf 70}, 045016 (2004) [hep-ph/0403296].

\bibitem{bowen} R. Bowen et al., Mon.\ Not.\ Roy.\ Astron.\ Soc.\ {\bf 334}, 760 (2002) [astro-ph/0110636].

\bibitem{QE}
W.~Hu and T.~Okamoto,
Astrophys.\ J.\  {\bf 574}, 566 (2002) [astro-ph/0111606].

\bibitem{trotta} R. Trotta and S.H. Hansen, Phys.\ Rev.\ D {\bf 69}, 023509 (2004) [astro-ph/0306588].

\bibitem{katrin}
A.~Osipowicz et al.\  [KATRIN Coll.],
hep-ex/0109033;
B.~Bornschein  [KATRIN Coll.],
contribution to the Proceedings of AHEP 2003, Valencia (Spain),
published in JHEP Proc.\ AHEP2003/064.

\bibitem{doublebeta}
S.R.~Elliott and P.~Vogel,
Ann.\ Rev.\ Nucl.\ Part.\ Sci.\  {\bf 52}, 115 (2002) [hep-ph/0202264].

\end{thebibliography}
\end{document}